# Ambient Listening in Clinical Practice: Evaluating EPIC Signal Data Before and After Implementation and Its Impact on Physician Workload

Yawen GUO[a,1], Di HU[a], Jiayuan WANG[a], Kai ZHENG[a], Danielle PERRET[b], Deepti PANDITA[c], Steven TAM[c]
[a]*Department of Informatics, University of California, Irvine, CA, USA*
[b]*Department of Physical Medicine & Rehabilitation, Irvine, CA, USA*
[c]*Department of Medicine, University of California, Irvine, CA, USA*
ORCiD ID: Yawen Guo https://orcid.org/0009-0006-2048-1798,
Di Hu https://orcid.org/0000-0002-2842-1478,
Kai Zheng https://orcid.org/0000-0003-4121-4948

**Abstract.** The widespread adoption of EHRs following the HITECH Act has increased the clinician documentation burden, contributing to burnout. Emerging technologies, such as ambient listening tools powered by generative AI, offer real-time, scribe-like documentation capabilities to reduce physician workload. This study evaluates the impact of ambient listening tools implemented at UCI Health by analyzing EPIC Signal data to assess changes in note length and time spent on notes. Results show significant reductions in note-taking time and an increase in note length, particularly during the first-month post-implementation. Findings highlight the potential of AI-powered documentation tools to improve clinical efficiency. Future research should explore adoption barriers, long-term trends, and user experiences to enhance the scalability and sustainability of ambient listening technology in clinical practice.

## 1. Introduction

The Health Information Technology for Economic and Clinical Health (HITECH) Act drove widespread Electronic Health Records (EHRs), which has increased documentation burden and contributed to clinician burnout[1,2]. In response, early efforts to reduce documentation workload have focused on technologies such as digital scribes and voice dictation. Most approaches targeted specific NLP tasks, including entity extraction, classification, and summarization of conversations[3].

The integration of large language models (LLMs) has enhanced ambient listening technology, enabling it to generate more detailed, accurate, and human-scribe-comparable documentation[4]. Multiple large health systems and academic institutions have conducted pilot studies to assess the effectiveness and usability of ambient listening tools and have shown their potential to improve physician-patient interactions. Haberle et al. evaluated physician engagement with an ambient listening tool using surveys, highlighting its potential to enhance physician experiences and the importance of efficient training[5]. Shah et al. conducted pre- and post-use paired surveys to assess user experience, demonstrating reduced physician task load, burnout, and improved usability[6]. Additionally, The Permanente Medical Group deployed ambient AI scribe technology for 10,000 physicians, supporting 303,266 patient encounters over 10 weeks, showing high-quality documentation, reduced clerical time, and positive feedback from both physicians and patients[7].

---
[1] Corresponding Author: Yawen Guo, yaweng5@uci.edu.

Despite promising early findings, most existing studies rely on surveys to examine user perspectives and experiences, with a limited focus on quantitative evaluation of system usage patterns and performance metrics[8,9]. This exploratory study is part of an ongoing quality improvement pilot project to evaluate the effectiveness of ambient listening tools across an integrated health system—The University of California, Irvine (UCI) Health, Orange, California—in outpatient settings, including primary care and various specialties. We present preliminary results from analyzing EPIC Signal data to evaluate the effectiveness of ambient listening tools on physician workload. We hypothesize that the use of the tool will affect note length and reduce clinician workload in terms of time spent on note-taking. Building upon prior work, this study aims to quantitatively assess adoption trends and signal metrics to evaluate the broader impact of ambient AI scribes on clinical efficiency, documentation quality, and physician workload, providing insights into their scalability and sustainability in real-world practice.

## 2. Method

In 2023, UCI Health launched a pilot project recruiting physicians to implement two commercially available ambient listening tools: DAX Copilot (Nuance Communications, Inc., Burlington, MA, USA) and Abridge (Abridge AI, Inc., Pittsburgh, PA, USA) in ambulatory care settings[10, 11]. This preliminary analysis included pilot participants who were enrolled for at least four weeks and generated a minimum of 10 notes using ambient listening tools.

Paired two-sample t-tests were conducted to compare the means of the following 4 Epic Signal metrics: average number of characters in a progress note, average number of characters documented per appointment (considering all type of notes generated in an appointment), average time spent writing a note, and time spent writing notes per day. The baseline represents the average values of these metrics recorded during a 3-month period immediately preceding the start of the ambient listening tool usage. The analysis also includes a 3-month follow-up period. We compared these metrics at baseline with their cumulative averages at first, second, and third months post-ambient listening pilot. The assumption of normality was assessed using the Shapiro-Wilk test. Outliers were removed using the Interquartile Range Method to ensure the normality assumption was not violated. T-tests were applied when data met normality ($p > 0.05$).

## 3. Results

A total of 124 physicians across different specialties enrolled in the ambient listening tool pilot, with the majority in primary care (55) and others in specialty care (61) and medical learners (8). Specialties range from gastroenterology, geriatrics, and psychiatry to rheumatology, cardiology, sports medicine, etc. 68 enrollees met our analysis inclusion criteria, having used the tool for more than 4 weeks and at least 10 notes. After addressing missing values and outliers to ensure data availability across all 4 signals, our final analysis included 42 participants with 1-month of use, 32 participants with 2-month of use, and 31 participants with 3-month of use.

**Table 1.** Comparative analysis of EPIC Signal metrics pre- and post-pilot

| EPIC Signal Metrics | Baseline V.S. Post 1M (n=42) | | Baseline V.S. Post 2M (n=32) | | Baseline V.S. Post 3M (n=31) | |
| --- | --- | --- | --- | --- | --- | --- |
| | Baseline | 1-Month | Baseline | 2-Month | Baseline | 3-Month |
| **Average characters per note** | 7952.37 | 8481.94 * | 7989.96 | 8490.73 * | 7974.7 | 8511.33 * |
| **Average characters per appointment** | 8232.41 | 9300.70 *** | 8348.73 | 9365.66 *** | 8294.71 | 9252.99 *** |
| **Average time per note** | 6.58 | 5.47 *** | 6.46 | 5.39 *** | 6.46 | 5.45 *** |
| **Average time per provider per day** | 63.99 | 54.40 *** | 64.75 | 54.23 *** | 63.28 | 53.33 *** |

*Statistical significance:* * = p<0.05, ** = p<0.01, *** = p<0.001

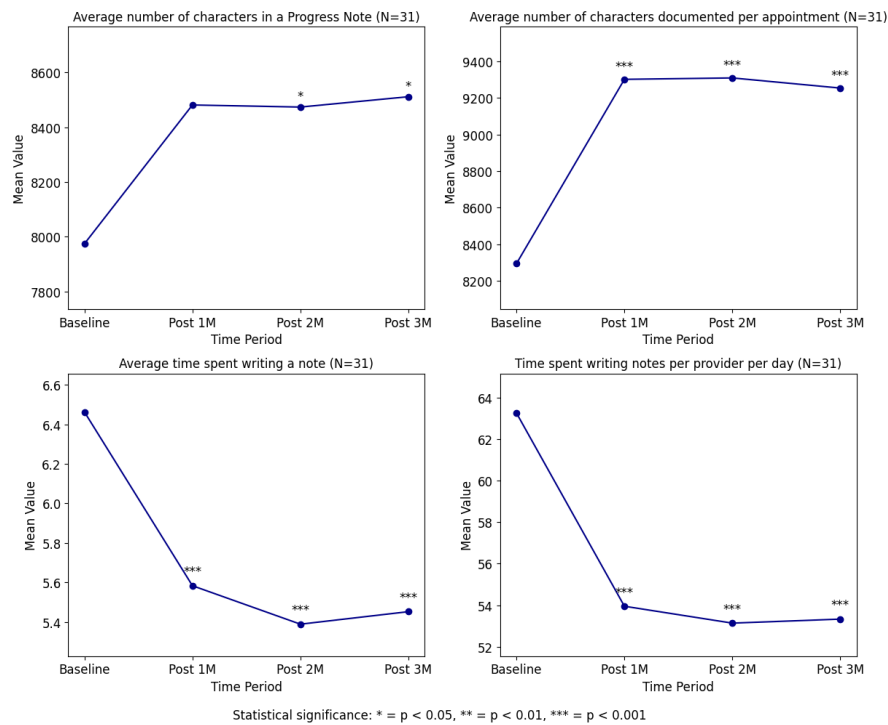

Statistical significance: * = p < 0.05, ** = p < 0.01, *** = p < 0.001

**Figure 1.** Longitudinal trends in documentation metrics following implementation

As shown in Table 1, documentation time significantly decreased following the implementation of the ambient listening tool pilot. The average time spent writing a note dropped from 6.58 minutes at baseline (during the 3 months prior to the pilot) to 5.45 minutes over the 3-months post-pilot (T-Statistic -6.61, $p<0.001$). Similarly, the average time spent on documentation per provider per day decreased from 63.99 minutes at baseline to 53.33 minutes during the post pilot period (T-Statistic -3.98, $p<0.001$). In contrast, note length increased following the pilot. The average number of characters per note increased by approximately 559 comparing the 3 months pre- and post-pilot (T-Statistic 2.64, $p<0.05$). Additionally, the average number of characters per appointment rose by about 1021, from 8232.41 at baseline to 9252.99 during the post-pilot period (T-Statistic 5.23, $p<0.001$). Figure 1 illustrates trends in Epic Signal metrics over the 3-month period, showing significant changes immediately after the initiation of ambient listening tool pilot, followed by smaller, continuous differences in subsequent months.

## 4. Discussion

Existing clinical informatics literature explores user experiences and perspectives on ambient listening tool usage[5-7], as well as its impact on EHR documentation time and deficiency rates[7-9]. Building upon this body of research, our study leverages system-logged EPIC Signal data to evaluate the effectiveness and utility of this tool, with a specific emphasis on clinician documentation workload and burden. Our findings from this preliminary analysis indicate that the use of ambient listening tools significantly reduced time spent on writing notes while increasing note length. Specifically, physicians were able to produce longer notes in less time, suggesting a potential for improving documentation efficiency and comprehensiveness. However, the increase in note length may also reflect the templates embedded in the ambient listening tools, which may include non-clinical and verbose content. Future studies should carefully review these templates to ensure a balance between documentation comprehensiveness and readability. Additionally, the first month of pilot witnessed the most significant changes in both time spent on notes and note length. This pattern may reflect a learning curve, where clinicians rapidly adapt to and optimize their use of the tool during the initial phase. The notable early gains suggest that focused training programs during the first month of implementation could facilitate smoother adoption and maximize the tools' benefits.

While 124 physicians enrolled in the pilot, only 46 (37%) were included in our analysis due to limited usage and the availability of corresponding Epic Signal data. Future research should investigate factors influencing differential adoption, including user-specific barriers, preferences, and perceptions of usefulness. Qualitative studies, such as interviews or surveys, may provide insights into these variations and inform strategies to enhance adoption and sustained use. Furthermore, although two different tools were rolled out during the pilot, this preliminary analysis focused solely on the general impact of ambient listening technology without conducting a comparative analysis between the tools. Future studies should incorporate larger sample sizes, longer follow-up periods, and comparisons of different products to further assess usability, long-term effectiveness, scalability, and impact of ambient listening on clinician satisfaction and patient outcomes.

## 5. Conclusion

Our preliminary results demonstrate that the use of ambient listening tools is statistically associated with reduced time spent on note-taking and increased note length. These findings suggest that while ambient listening tools may alleviate documentation workload, they could also result in more lengthy notes for physicians to review. Additionally, we found that the most significant reduction in note-taking time occurred during the first month after pilot implementation, indicating the presence of a learning curve that may benefit from targeted training. Future research should explore adoption barriers, long-term usage trends, and user experiences to guide continued optimization and implementation strategies for ambient documentation tools. Studies incorporating larger cohorts and qualitative assessments are needed to further elucidate the broader impacts of these tools on clinical practice and patient care.